\documentclass[12pt,preprint,epsf]{aastex}
\slugcomment{\today}

\newcommand{\G}{\Gamma}

\newcommand{\sT}{\sigma_{\rm T}}

\newcommand{\e}{\epsilon}
\newcommand{\g}{\gamma}
\newcommand{\gp}{\gamma^{\prime}}

\newcommand{\dD}{\delta_{\rm D}}

\newcommand{\psim}{\lower.5ex\hbox{$\; \buildrel \propto \over\sim \;$}}
\newcommand{\lbar}{\lower.0ex\hbox{$\; \buildrel
{\lower0.0ex \hbox{-}} \over\lambda  \;$}}

\newcommand{\erg}{\mathrm{erg}}

\newcommand{\s}{\mathrm{s}}

\newcommand{\pc}{\mathrm{pc}}

\shorttitle{Spectral Break in 3C~454.3}
\shortauthors{Finke \& Dermer}

\begin{document}
\title{On the Break in the {\em Fermi}-LAT Spectrum of 
3C~454.3 }

\author{Justin D. Finke$^{1}$ and Charles D. Dermer}

\affil{U.S.\ Naval Research Laboratory, Code 7653, 4555 Overlook Ave SW,
       Washington, DC, 20375-5352\\
$^1$NRL/NRC Research Associate}
     
\email{justin.finke@nrl.navy.mil}

\begin{abstract}
{\em Fermi} Gamma ray Space Telescope observations of the flat
spectrum radio quasar 3C~454.3 show a spectral-index change $\Delta
\Gamma \cong 1.2\pm 0.3$ at break energy $E_{br} \approx 2.4\pm0.3$
GeV. Such a sharp break is inconsistent with a cooling electron
distribution and is poorly fit with a synchrotron self-Compton model. 
We show that a combination of two
components, namely the Compton-scattered disk and broad-line region
(BLR) radiations, explains this spectral break and gives a good fit to
the quasi-simultaneous radio, optical/UV, X-ray, and $\gamma$-ray
spectral energy distribution observed in 2008 August.
%for the source suggest that this 
%is the most likely explanation for this break.
A sharp break can be produced independent of the emitting region's
distance from the central black hole if the BLR has a gradient in
density $\propto R^{-2}$, consistent with a wind model for the BLR.
\end{abstract}

\keywords{gamma rays:  galaxies --- Quasar:  3C~454.3 --- 
radiation mechanisms:  non-thermal
}

\section{Introduction}
\label{intro}

%3C454.3 underwent a giant outburst in 2007 observed with
%the $\g$-ray observatory {\em AGILE} \citep{vercellone08} as well as
%the {\em Swift} and {\em INTEGRAL} satellites, and the many
%ground-based optical and radio telescopes in the Whole Earth Blazar
%Telescope (WEBT) consortium \citep{raiteri08,vercellone09}.  

The flat spectrum radio quasar (FSRQ) \object{3C~454.3}
(\object{PKS~2251+158}) is representative of blazars with large,
$\gtrsim 10^{48}$ erg s$^{-1}$ apparent $\gamma$-ray luminosities, and
is one of the brightest sources seen with the recently launched {\em
Fermi} Gamma-Ray Space Telescope
\citep{tosti08_atel,escande09_atel}. It has also been detected with
AGILE up to $\approx 3$ GeV, including a giant outburst in 2007 prior
to the {\em Fermi} launch \citep{vercellone09,vercellone10}.  Based on
$> 100$ MeV data taken during the checkout and early science phase of
the mission (2008 July 7-- October 6), this source was found to be
variable on timescales down to $t_{v}\sim$ few days
\citep{abdo09_3c454.3}.  A well-sampled data set using the
ground-based SMARTS (Small and Moderate Aperture Research Telescope
System) in the B, V, R, J, and K bands, and the {\em Swift} X-ray
telescope (XRT) and ultraviolet-optical telescope (UVOT) was collected
simultaneously with {\em Fermi} observations in the period 2008 August
through December \citep{bonning09}.  These multiwavelength
observations revealed that the infrared, optical, and $\g$-rays were
have similar variability patterns.  The $\gamma$ rays display similar
relative flux changes compared to the longer optical wavelengths, but
smaller relative flux changes as the short wavelength optical
emission, which \citet{bonning09} interpreted as being due to an
underlying accretion disk.  The {\em Swift} X-ray data did not
strongly correlate with the other wavelengths, though
\citet{jorstad10} and \citet{bonnoli10} find a correlation at
different epochs.

The {\em Fermi} observations of \object{3C~454.3} revealed a broken
power-law spectrum, with $\Gamma\cong 2.3\pm 0.1$ (where the photon
flux $\Phi\propto E^{-\Gamma}$) and $\Gamma\cong 3.5\pm 0.25$, below
and above a break energy of $2.4\pm0.3$ GeV, respectively
\citep{abdo09_3c454.3}.  This spectrum is not consistent with Compton
scattering of a lower-energy photon source by jet electrons in the
fast- or slow-cooling regimes, the standard interpretation for the
high energy spectrum in leptonic models of blazars.
\citet{abdo09_3c454.3} also consider it unlikely that the spectral
break is a result of $\g\g$ absorption with a local radiation field or
the extragalactic background light (EBL), the former because this
would require a lower jet bulk Lorentz factor than that inferred from
superluminal radio observations of \object{3C~454.3}
\citep[$\Gamma_{bulk}\approx15$;][]{jorstad05}, and the latter because
the universe is transparent to 40~GeV photons at $z\la1.0$ for all EBL
models \citep[e.g.,][]{finke10_EBLmodel,stecker06,gilmore09}.
\citet{abdo09_3c454.3} thus consider the most likely explanation to be
that the spectral break is caused by an intrinsic break in the
electron distribution.  However, as explained below, this is unlikely,
given the broadband spectral energy distribution of this source.
%The cause of this break remains a puzzle.

In this letter, we show that a combination of two Compton scattering
components can explain the break in the LAT spectrum of
\object{3C~454.3}. In particular, we consider a combination of
components from Compton-scattered accretion disk and broad-line region
(BLR) radiation, both of which should be strong seed photon sources in
FSRQs.  The implications of this model for the BLR and for the
variability of the emitted radiation are explored.  Finally, we
conclude with a brief discussion, with implications for breaks in
other sources.

We use parameters $H_0=71$ km s$^{-1}$ Mpc$^{-3}$,
$\Omega_m=0.27$, and $\Omega_\Lambda=0.73$ so that \object{3C~454.3}, 
with a redshift of $z=0.859$, has a luminosity distance of 
$d_L=5.5$ Gpc. 

\section{The Spectral Energy Distribution of 3C~454.3}

\begin{figure}
%\epsscale{1.2}
\epsscale{1.1}
\plotone{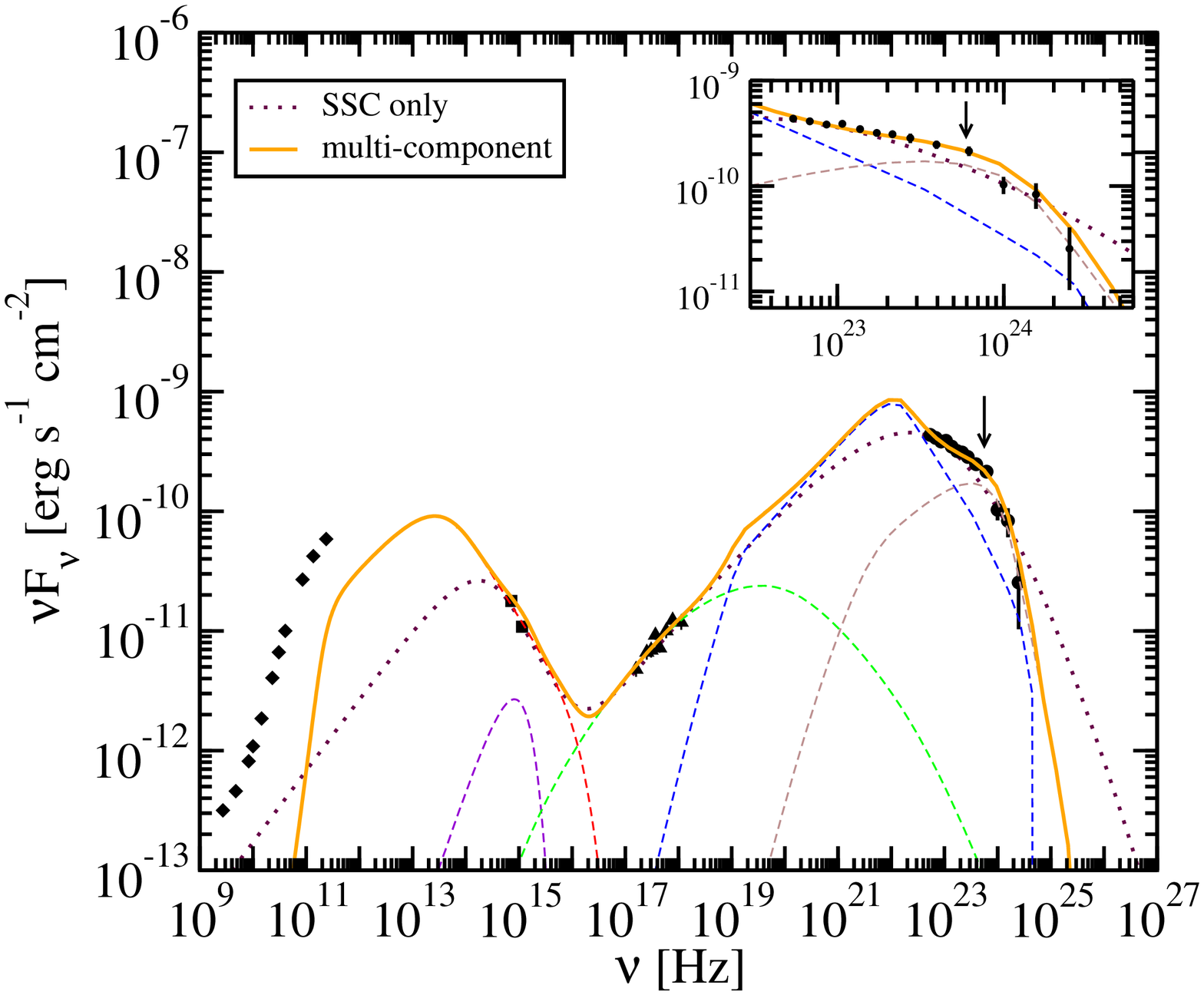}
\caption{ The SED of 3C~454.3.  The Effelsberg/IRAM radio data are
given by the diamonds, the {\em Swift}-UVOT and XRT data are given by
the squares and triangles, respectively, and the {\em Fermi}-LAT data
by the filled circles.  The model components for the multi-component
model are shown as dashed curves, where the purple dashed curve is the
Shakura-Sunyaev disk, the red is the synchrotron, the green is the
SSC, the blue is the Compton-scattered disk, and the brown is the
Compton-scattered BLR component assuming $\tau_{BLR}=0.01$.  The
thick dotted curve gives the SSC fit, and the thick solid curve shows
the multi-component model.  The inset shows the {\em Fermi}-LAT
spectrum in more detail.  The arrows indicate the frequency of the
spectral break. }
\label{3c454.3sed}
\end{figure}

The contemporaneous SED data from \object{3C~454.3} are given in
\citet{abdo09_3c454.3} and plotted in Fig.\ \ref{3c454.3sed}.  
The radio data are likely from a larger region of the jet than the
rest of the SED, and can be treated as upper limits for the purposes
of model fits.  We assume that the electron distribution which
produces this radiation has the form of a broken power-law given by
$n_e \propto \g^{\prime -p_1}$ for $\gp_{min} < \gp<\gp_{brk}$ and
$n_e \propto \g^{\prime -p_2}$ for $\gp_{brk} < \gp < \gp_{max}$.  The
{\em Swift} UVOT data give a power-law spectrum with
$\Gamma_{opt}\approx2.9$, presumably from electron synchrotron
emission.  The X-rays from the {\em Swift} XRT give
$\Gamma_X\approx1.4$, which should be the index of the synchrotron
emission below the break, assuming that the X-rays are from
Compton-scattering with the same electrons as are causing the
lower-energy synchrotron emission.  Thus the {\em Swift} data imply
that, if the electron distribution can indeed be represented by a
double power-law, $p_1\approx1.8$, and $p_2\approx4.8$, since
$p=2\G-1$.  This spectrum indicates a fast cooling regime, where
one expects $p_1\approx 2$ and $p_2=q+1$, where $q$ is the injection
index.

If the LAT spectrum was the result of Thomson scattering of a single
source of photons, whether that source is the disk, BLR, or dust
torus, the spectral indices would be about the same as for the
synchrotron, namely, $\Gamma\approx1.4$ and $\Gamma\approx2.9$ well
below and above the break, respectively.  Instead, $\Gamma \approx
2.3$ and $\Gamma \approx 3.5$ are observed, respectively, although it
is possible that Klein-Nishina (KN) effects could soften the higher energy
index. The lower energy LAT index furthermore does not necessarily
correspond to the asymptotic value at $E\approx E_{br}$.

\citet{abdo09_3c454.3} suggested that the break could be caused by the
high energy cutoff in the electron distribution, $\g_{max}$.  However,
if this is the case, below the $\g$-ray break, one would expect the
same $\G$ as in the optical ($\G_{opt}=2.9$), because of their
correlated variability, instead of the $\G=2.3$ which is observed.
The $\g$-ray spectral index could be softer than $\G_{opt}$, due to
KN effects, but it is unlikely to be harder.

It is possible that the X-rays through {\em Fermi} $\g$-rays are all
created by a single component, likely the SSC component, since this is
generally broader than external Compton components.  In this model,
the softer $\g$-rays are created by the portion of the SSC component
at and below the break, as it gradually softens.  This SSC-only model
gives an adequate fit to the broadband SED, although it  gives 
a poor fit to the LAT data (Fig.\ \ref{3c454.3sed}), and does not
produce a sharp break in the LAT $\g$-ray spectrum, although
\citet{abdo09_3c454.3} found that the LAT spectrum could just as
easily be fit by a smooth transition such as, for example, a power-law
with an exponential cutoff.  See Table~1 for the spectral parameters
for this fit and \citet{finke08_SSC} for details on this model.  The
SSC-only model in Fig.\ \ref{3c454.3sed} does not enter the KN regime
until around $\nu\ga3\times10^{26}$ Hz, which is considerably above
the {\em Fermi}-LAT data, and these $\g$-rays are thus produced
entirely in the Thomson regime.  However, it is possible that local
$\g\g$ absorption could soften the higher energy portion of the
spectrum and produce a better fit.

%\clearpage

\begin{deluxetable}{ccc}
\label{modelparams1}
%\rotate
\tabletypesize{\scriptsize}
\tablecaption{
Model Parameters\tablenotemark{1}.
}
\tablewidth{0pt}
\tablehead{
\colhead{Symbol} &
\colhead{Multi-Component model} &
\colhead{SSC model}
}
\startdata
$z$		& 0.859	& 0.859  \\
$\G_{bulk}$	& 15	& 15  \\
$\dD$       & 30    	& 15 \\
$B$         & 0.4 G   	& 0.032 G \\
$t_v$       & $1.7 \times 10^5$ s & $1.7\times10^5$ s \\
\hline
$p_1$       & 2.0 & 1.8    \\
$p_2$       & 4.4 & 4.8	 \\
$\gp_{min}$  & $3\times10^1$ & 10 \\
$\gp_{max}$  & $2\times 10^4$ & $2\times10^9$ \\
$\gp_{brk}$ & $1.1\times 10^3$ & $10^4$ \\ 
\hline
$M_9$       & 2.0   \\
$l_{Edd}$   & 0.04     \\
$\eta$ & 1/12 \\
$r$ & $1.5\times 10^3 R_g$ \\
\hline
$\tau_{BLR}$ & 0.01 \\
$R_i$ & $5.0\times10^2 R_g$ \\
$R_o$ & $5.0\times10^5 R_g$ \\
$\zeta$ & $-2$ \\ 
\hline
$B/B_{eq}$ & 0.6 & 0.06 \\
$P_{j,B}$ & $1.8\times10^{45}$ erg s$^{-1}$ & $10^{43}$ erg s$^{-1}$ \\
$P_{j,par}$ & $2.7\times 10^{46}$ erg s$^{-1}$ & $2.3\times10^{47}$ erg s$^{-1}$ \\
\enddata
\tablenotetext{1}{See text and \citet{dermer09}
for details on parameters and calculations of jet powers.}
\end{deluxetable}

But the SSC model has other problems.  It ignores the BLR for this
source, which must be present due to the strong broad emission lines
seen in this object's optical spectrum; it results in an extremely low
magnetic field, $B$, which is considerably lower than the
equipartition value, $B_{eq}$, and thus the particle jet power,
$P_{j,par}$ is much greater than the magnetic field jet power,
$P_{j,e}$ and close to the Eddington limit (see Table 1); and
it may not be consistent with the variability of this source (see \S\
\ref{variability} below).  The constraints on the Doppler factor
from \citet{jorstad05} and the constraints on the emitting region size
from the variability timescale imply $B = 0.032\pm0.005$ G for the SSC
model.  We propose that the LAT spectrum of \object{3C~454.3} is
instead the result of a combination of two Compton-scattered
components, for example, Compton scattering of the disk and BLR
radiation.

\section{BLR Radiation Energy Density}

Before modeling the SED of 3C 454.3, we derive approximate formulae for the energy
density of the BLR at the location of the jet. A regime where the energy densities of the
Compton-scattered disk and BLR components are approximately equal 
can explain the LAT spectrum of \object{3C~454.3}.

We assume that the BLR is spherically symmetric and extends from inner
radius $R_i$ to outer radius $R_o$, and the electron number density of
the BLR is a function of distance from the central source, $n_{BLR}(R)
= n_0(R/R_i)^\zeta$, so that the Thomson depth of the BLR is given by
$\tau_{BLR} = \sT \int_{R_i}^{R_o} dR\ n_{BLR}(R)$ giving, when $R_i \ll R_o$,
\begin{eqnarray}
\label{tauBLR}
%\\ %\nonumber
\tau_{BLR} \cong \sT n_0\ \left\{ \begin{array}{ll}
	\frac{R_o}{\zeta+1}\left( \frac{R_o}{R_i} \right)^\zeta\; ,
	& \zeta>-1 \\
	\frac{-R_i}{\zeta+1}\; , & \zeta<-1 
	\end{array}
	\right.\ .
\end{eqnarray}

The BLR Thomson-scatters photons from the accretion disk, assumed 
to be a point source located at origin with luminosity $L_d$.  
If the distance from the black hole to the $\g$-ray emitting region, 
$r<R_i$, the energy density from BLR radiation is approximately independent of $r$, so
\begin{eqnarray}
u_{BLR}(r<R_i) \cong \frac{L_d \sT}{4\pi c} \int_{R_i}^{R_0} \frac{dR}{R^2}\ n_{BLR}(R)
\nonumber
\\ 
\cong \frac{L_d \tau_{BLR}}{4\pi c}\left\{  \begin{array}{ll}
	\frac{1}{R_i^2}\frac{\zeta+1}{\zeta-1}\; , & \zeta<-1 \\
	\frac{1}{R_iR_o}\left( \frac{R_i}{R_o}\right)^\zeta 
	\frac{\zeta+1}{\zeta-1}\; ,
	& -1<\zeta<1 \\
	\frac{1}{R_0^2}\frac{\zeta+1}{\zeta-1}\; , & \zeta>1
	\end{array}
	\right.\ ,
\end{eqnarray}
using eqn.\ (\ref{tauBLR}).
If $R_i<r<R_o$, then, assuming the logarithm term in eqn.\ (93) of 
\citet{dermer09} is $\approx 1$, 
$$
u_{BLR}(r) \cong \frac{L_d \sT}{4\pi c r} \int_{R_i}^{R_0} 
\frac{dR}{R}\ n_{BLR}(R)\ 
$$
\begin{equation} \cong\frac{L_d \tau_{BLR}}{4\pi c}\left\{  \begin{array}{ll}
	\frac{1}{rR_i}\left(\frac{r}{R_i}\right)^\zeta 
	\frac{\zeta+1}{\zeta}\; ,
	& \zeta<-1 \\
	\frac{1}{rR_o}\left(\frac{r}{R_o}\right)^\zeta 
	\frac{\zeta+1}{\zeta}\; ,
	& -1<\zeta<0 \\	
	\frac{1}{R_iR_o}\left(\frac{R_i}{R_o}\right)^\zeta
	\frac{\zeta+1}{\zeta-1}\; ,
	& 0<\zeta<1 \\	
	\frac{1}{R_o^2}\ \frac{\zeta+1}{\zeta-1}\; ,
	& \zeta>1 
	\end{array}
	\right.\ .
\end{equation}
\citep[see also][]{donea03}.

\section{Compton Scattering Disk and BLR Radiation}

The external Compton-scattered BLR 
$\nu F_{\nu}$ flux in the Thomson regime can be approximated as 
\begin{equation}
\label{ECBLRflux}
f_\e^{ECBLR} = \frac{\dD^6}{6\pi d_L^2} c\sT 
	u_{BLR}
	\tilde\g^{\prime 3} N_e(\tilde\gp)
\end{equation}
where
$
\tilde\gp = \sqrt{\e(1+z)/\left[\dD^2\e_d(8.4r_g)\right]}
$
%$$
%\tilde\gp = \frac{1}{\dD}\sqrt{\frac{\e(1+z)}{\e_d(8.4r_g)}}\ 
%$$
\citep{dermer94} and $\e$ is the observed photon energy.  For
\object{3C~454.3}, $\G_{bulk}\approx 15$, the Doppler factor
$\delta_D\approx25$ \citep{jorstad05}.  \citet{gu01} found $M_9 =
M_{BH}/(10^9 M_\odot) \approx 4.4$ where $M_{BH}$ is the black hole
mass.  A more recent analysis of the black hole mass gives $M_9=1-3$
\citep{bonnoli10}, and we assume for the remainder of this work that
$M_9=2$.  If the $\g$-ray emission region is located at $r \ll
\G_{bulk}^4 r_g \approx 10^2$ pc, where
$r_g=GM_{BH}c^{-2}=1.5\times10^{14}M_9$ cm, then the external
Compton-scattered disk emission is in the near field regime.  In this
case the received $\nu F_\nu$ flux can be approximated by
\begin{equation}
\label{ECdiskNFflux}
f_\e^{ECD} = \frac{\dD^6}{2}\frac{c\sT}{6\pi d_L^2} 
	\left[ \frac{L_d r_g}{4\pi r^3 c}\right]
	\bar\g^{\prime 3} N_e(\bar\gp)\;,
\end{equation}
where
$
\bar\gp = 2\sqrt{(1+z)\e/\left[\dD^2\e_d(\sqrt{3}r)\right] }\ 
$
%$$
%\bar\gp = \frac{2}{\dD}\sqrt{\frac{(1+z)\e}{\e_d(\sqrt{3}r)} }\ 
%$$
\citep{dermer02}. Here
$$
\e_d(R)=1.5\times 10^{-4} \left[ \frac{l_{Edd}}{M_9\eta}\right]^{1/4}
\left(\frac{R}{r_g}\right)^{-3/4} 
$$
is the mean photon energy from a \citet{shakura73} accretion disk at
$R$, $l_{Edd}=L_d/L_{Edd}$, $L_{Edd}=1.3\times10^{47}M_9$, and $\eta$
is the accretion efficiency.

A solution where the Compton-scattered disk and BLR emission are
approximately equal can explain the {\em Fermi}-LAT emission.
Also, a robust solution is obtained by noting that both
$f^{ECBLR}_\e$ and $f^{ECD}_\e$ are $\propto r^{-3}$ if $\zeta = -2$
and $R_i<r<R_o$.  A BLR gradient of $\zeta=-2$ is consistent with a
wind model \citep{murray95,chiang96}.  Setting eqns.\
(\ref{ECBLRflux}) and (\ref{ECdiskNFflux}) equal to each other, and,
ignoring the fact that the electrons generally scatter photons of
different energies, one obtains the condition $\tau_{BLR} = (R_i/r_g)^{-1}$
%\begin{equation}
%\label{ECradius}
%\tau_{BLR} = \left( \frac{R_i}{r_g}\right)^{-1}\; 
%\end{equation}
%For parameters appropriate for \object{3C~454.3}, i.e., 
%$\dD=15$, $R_{BLR}= 3\times 10^{18}$ cm, $M_9=1$, and 
%$\tau_{BLR}=10^{-2}$, one gets $r=3.4\times 10^{17}$ cm.
for emission regions formed in the BLR. 
In order to have comparable flux from the two 
components, this 
relation shows that  the total Thomson 
depth through the wind should increase in inverse
proportion to the inner radius of the BLR.

An accurate fit to the multiwavelength SED
using the multi-component model of
\citet{dermer09}, using model
parameters from Table~1 with $\tau_{BLR} = 0.01$, 
is seen in Fig.\ \ref{3c454.3sed}.  
%For details on these parameters, which
%are consistent with inferences from observations of \object{3C~454.3},
%see \citet{dermer09}.  
In this model, the portion of the {\em Fermi}
spectrum below the break is from Compton scattering of a combination
of direct accretion disk and BLR-reprocessed radiation, while 
above the break it is almost entirely from Compton scattering of
BLR-reprocessed disk photons.  This model uses the full Compton 
cross section for all Compton calculations, and the full
Shakura-Sunyaev disk spectrum for both direct and BLR-reprocessed
Compton-scattered emission.  For the Compton-scattered BLR
calculation, the Shakura-Sunyaev disk spectrum is assumed to originate
from the black hole itself, whereas in the direct disk-scattered
calculation, geometric effects are taken into account.

For KN effects to become important in Compton scattering
the BLR radiation, $\g^\prime \ga (4\G_{bulk}\e_d)^{-1}$ where $\e_d$
is the typical dimensionless energy of photons from the disk in the
disk's stationary frame.  For our model, $\e_d= 2.2\times 10^{-5}$,
and $\G_{bulk}=15$, so that $\g^{\prime}_{KN}=760$.  Electrons with
this energy scatter photons primarily to energies of $ \e_{KN} \approx
\dD^2 \e_d\g^{\prime 2}_{KN}/(1+z) = 6.2\times 10^3 $
%$$
%\e_{KN} \approx \frac{\dD^2 \e_d}{1+z}\g^{\prime 2}_{KN} = 1.4\times 10^3
%$$
which corresponds to frequencies of $\nu = 7.6\times 10^{23}$ Hz.
Thus above the break the Compton scattering of the
BLR is almost entirely in the KN regime.  Since the
KN cross section goes as approximately
$\sigma_{KN}\propto\e^{-1}$, $f^{BLR}_\e$ should go as $\G\approx
\G_{opt}+1 = 3.9$ above the break in the $\g$-ray spectrum.  This is
quite close to the $\G=3.5$ which is observed by {\em Fermi}.
Although the BLR photons are Thomson-scattered disk photons with
essentially the same energy as the direct disk photons, geometric
effects from Doppler boosting and aberration make Compton-scattered
BLR radiation peak at higher photon energies than the disk photons,
which tend to come from behind.

The threshhold for $\g\g$ absorption from the
BLR-reprocessed disk radiation implies that $\g$-rays with energies
$
\e \ga 2[\e_d(8.4r_g)(1+z)]^{-1}
$
%$$
%\e \ga \frac{2}{\e_d(8.4r_g)(1+z)}
%$$
will be absorbed.  For our model 
this turns out to be $\e\approx 4.9\times 10^4$ or
$\nu\approx 6.0\times 10^{24}$ Hz, which is above the highest energy
photon observed with the {\em Fermi}-LAT from this source.

The sum of the particle and magnetic-field power $P_{j,par}+P_{j,B}$
is well below $L_{Edd}=2.6\times10^{47}\ \erg\ \s^{-1}$ for $M_9=2.0$
(the photon power is $\approx 8\times 10^{46}$ erg s$^{-1}$).
The model is close to equipartition between the total
particle and magnetic-field energy densities, assuming a factor
$10 \times$ more energy in total nonthermal particles than electrons,
i.e., $P_{j,par} = 10P_{j,e}$. Our multi-component fit has lower jet
power   than the fit to the same SED
by \citet{ghisellini09}, due primarily to their assumption of
a larger proton power, with $P_{j,par}>200\times
P_{j,e}$.  
%As a result, the ratio $P_{j}/L_{disk}$ for our fit is
%$<1$, putting it in a region of parameter space not well-occupied by
%fits to the brighest {\em Fermi}-LAT blazars in Fig.\ 12 of
%\citet{ghisellini09}.

\section{Variability}
\label{variability}

The $\g$-ray spectral break observed with {\em Fermi} was constant
over several month timescales \citep{abdo09_3c454.3}.  As the
$\g$-ray-emitting region moves outward, the spectral break should
remain at approximately the same energy, since for
$\zeta=-2$, the BLR- and disk-scattered flux should remain
approximately equal.  The overall $\g$-ray flux will, of course,
change over time, but as long as $R_i<r\ll \G^4r_g$, and $R_i$ or $\tau_{BLR}$
does not change significantly, the $\g$-ray spectral break should be
present.

In \citet{bonning09}, optical observations are seen to vary with
changes on the same magnitude as the $\g$-rays.  If the $\g$-rays were
produced by SSC emission, ``quadratic variability'', i.e.,
$f_\e^{SSC}\propto (f_\e^{syn})^2$, might be expected.  The absence of
such variability is in accord with our spectral fitting results means
that this model for 3C 454.3 is disfavored.

\begin{figure}
\epsscale{1.0}
\plotone{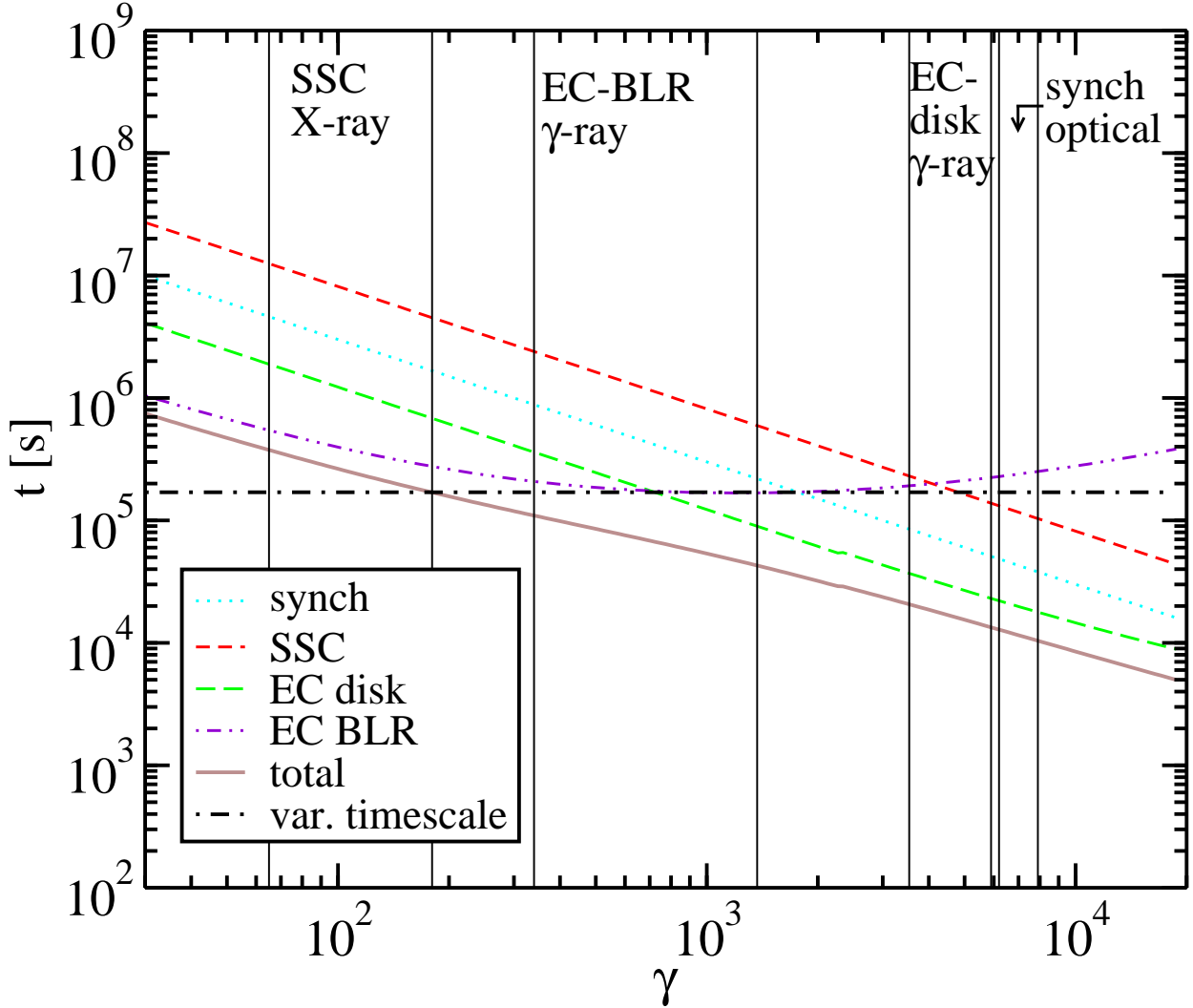}
\caption{ Cooling timescales in the observer's frame. The
electrons with $\g$ which produce the majority of the radiation in
certain wavebands are marked.  The electrons which produce X-rays 
from SSC emission have total cooling timescales longer than the 
variability timescale, explaining the lack of X-ray variability. 
The optical synchrotron radiation is made by electrons with
$\gamma\approx$ 6--8$\times10^{3}$, which would therefore correlate with the 
$\gamma$ rays formed by Compton-scattered disk radiation.}
\label{cooltime}
\end{figure}

The X-ray light curve of \object{3C~454.3} shows little variability
during this time, and is not correlated with the IR, optical, UV, or
$\g$-rays \citep{bonning09}.  This can be explained by the longer
cooling timescale (see Fig.\ \ref{cooltime}).  These timescale
calculations use monochromatic approximations for the
Compton-scattered disk and BLR components, but the full
Compton cross section for all components.  The electrons which
make the X-rays by SSC have lower energies than those that make the
optical and $\g$-rays, and thus cool more slowly than the observed variability 
timescale.

%\clearpage
 
The lack of correlated variability between X-rays and optical and
$\g$-rays might indicate that the X-rays originate from another part
of the jet than the LAT $\g$-rays.  However, in this case, the
emission would still likely be in the fast cooling regime, since the
$\g$-ray flux is significantly greater than the optical and radio
fluxes, and as long as one is in the fast cooling regime, Compton
scattering of a single component cannot explain the LAT spectrum.
Furthermore, during the 2007 November flares of {\object 3C~454.3},
detected by {\em Swift} UVOT, XRT and BAT, {\em INTEGRAL}, and {\em
AGILE}, the X-rays were weakly variable, compared with other
wavebands, but did seem to show correlated variability
\citep{vercellone09}, which indicates they originate from the same
emission region.  One final possibility is that a hadronic model may
be possible to explain the LAT $\g$-ray spectrum
\citep[e.g.,][]{muecke01}, however, such models may have difficulties
in explaining the correlated variability seen in this source, since
protons would evolve on longer timescales than electrons.

\section{Discussion}

%We have shown that an SSC model is 
%unlikely to explain quasi-simultaneous observations of 3C 454.3. 
Assuming the dual-component Compton-scattering scenario presented in
this paper is correct, what can it tell us about the location of the
$\g$-ray emission region?  It implies that this region is $R_i < r \ll
\G^4r_g$, so that, using the parameters from the modeling of 3C~454.3,
the location of the jet falls over a large range from $0.1\ \pc < r \ll
100\ \pc$.  However, the external energy density from disk and BLR
radiations is $\propto r^{-3}$, a steep decline, so that the blob 
must be within $\approx 0.1$ pc for a significant scattering 
component. Recent modeling of 3C 454.3 \citep{bonnoli10} also finds 
that the $\gamma$-ray emitting region is within the BLR.  
By contrast, \citet{sikora08} suggested that the $\g$-rays
observed by EGRET and AGILE were caused by Compton-scattering of
radiation from hot dust by a blob located $\sim 10$ pc from the
central black hole, though detailed  modeling is still needed
to determine if this model can explain the sharp spectral break in the
{\em Fermi} observations.  The blazar 3C 279 also shows a similar
spectral break during the onset of an extended episode of optical
polarization, while the jet emission region could still be within the
BLR region \citep{abdo10_3c279}.

Whether this model can predict similar spectral breaks in other FSRQs
and low-synchrotron-peaked blazars, such as AO 0235+164 or PKS
1502+106 \citep{abdo10_sed}, depends mainly on the properties of the
BLR in these sources, in particular, whether $\tau_{BLR} \approx
r_g/R_i$.  In specific blazars, especially at higher $z$,
$\g\g$-absorption by scattered disk radiation \citep{reimer07} could
still be effective.  Detailed modeling of simultaneous SED data will
test these scenarios, as it does for the case of \object{3C~454.3}.
If our model is correct, it provides evidence for a wind model of the
BLR \citep{murray95,chiang96,elvis00}.

%The double-peaked emission lines from
%accretion disks could appear single-peaked due to interactions with a
%wind \citep{murray97,eracleous03}.  It does seem that broad absorption
%lines, a signature of this disk wind model, exist in a bright X-ray
%source if those X-rays originate from a jet \citep{ghosh08}.

% and \citet{flohic08} 
%have shown that these winds could explain double-peaked 
%emission lines seen in some AGN.  

\acknowledgements 
We are grateful to the anonymous referee for helpful comments.
J.D.F. was supported by NASA Swift Guest Investigator Grant
DPR-NNG05ED411 and NASA GLAST Science Investigation DPR-S-1563-Y.
C.D.D. was supported by the Office of Naval Research.

%!****************************************************


\begin{thebibliography}{28}
\expandafter\ifx\csname natexlab\endcsname\relax\def\natexlab#1{#1}\fi

\bibitem[{{Abdo} {et~al.}(2009)}]{abdo09_3c454.3}
{Abdo}, A.~A. {et~al.} 2009, \apj, 699, 817

\bibitem[{{Abdo} {et~al.}(2010{\natexlab{a}})}]{abdo10_3c279}
---. 2010{\natexlab{a}}, Nature, 463, 919

\bibitem[{{Abdo} {et~al.}(2010{\natexlab{b}})}]{abdo10_sed}
---. 2010{\natexlab{b}}, \apj, 710, 1271

\bibitem[{{Bonning} {et~al.}(2009){Bonning}, {Bailyn}, {Urry}, {Buxton},
  {Fossati}, {Maraschi}, {Coppi}, {Scalzo}, {Isler}, \& {Kaptur}}]{bonning09}
{Bonning}, E.~W., {Bailyn}, C., {Urry}, C.~M., {Buxton}, M., {Fossati}, G.,
  {Maraschi}, L., {Coppi}, P., {Scalzo}, R., {Isler}, J., \& {Kaptur}, A. 2009,
  \apjl, 697, L81

\bibitem[{{Bonnoli} {et~al.}(2010){Bonnoli}, {Ghisellini}, {Foschini},
  {Tavecchio}, \& {Ghirlanda}}]{bonnoli10}
{Bonnoli}, G., {Ghisellini}, G., {Foschini}, L., {Tavecchio}, F., \&
  {Ghirlanda}, G. 2010, \mnras, submitted, arXiv:1003.3476

\bibitem[{{Chiang} \& {Murray}(1996)}]{chiang96}
{Chiang}, J. \& {Murray}, N. 1996, \apj, 466, 704

\bibitem[{{Dermer} {et~al.}(2009){Dermer}, {Finke}, {Krug}, \&
  {B{\"o}ttcher}}]{dermer09}
{Dermer}, C.~D., {Finke}, J.~D., {Krug}, H., \& {B{\"o}ttcher}, M. 2009, \apj,
  692, 32

\bibitem[{{Dermer} \& {Schlickeiser}(1994)}]{dermer94}
{Dermer}, C.~D. \& {Schlickeiser}, R. 1994, \apjs, 90, 945

\bibitem[{{Dermer} \& {Schlickeiser}(2002)}]{dermer02}
---. 2002, \apj, 575, 667

\bibitem[{{Donea} \& {Protheroe}(2003)}]{donea03}
{Donea}, A. \& {Protheroe}, R.~J. 2003, Astroparticle Physics, 18, 377

\bibitem[{{Elvis}(2000)}]{elvis00}
{Elvis}, M. 2000, \apj, 545, 63

\bibitem[{{Escande} \& {Tanaka}(2009)}]{escande09_atel}
{Escande}, L. \& {Tanaka}, Y.~T. 2009, The Astronomer's Telegram, 2328, 1

\bibitem[{{Finke} {et~al.}(2008){Finke}, {Dermer}, \&
  {B{\"o}ttcher}}]{finke08_SSC}
{Finke}, J.~D., {Dermer}, C.~D., \& {B{\"o}ttcher}, M. 2008, \apj, 686, 181

\bibitem[{{Finke} {et~al.}(2010){Finke}, {Razzaque}, \&
  {Dermer}}]{finke10_EBLmodel}
{Finke}, J.~D., {Razzaque}, S., \& {Dermer}, C.~D. 2010, \apj, 712, 238

\bibitem[{{Ghisellini} {et~al.}(2009){Ghisellini}, {Tavecchio}, \&
  {Ghirlanda}}]{ghisellini09}
{Ghisellini}, G., {Tavecchio}, F., \& {Ghirlanda}, G. 2009, \mnras, 399, 2041

\bibitem[{{Gilmore} {et~al.}(2009){Gilmore}, {Madau}, {Primack}, {Somerville},
  \& {Haardt}}]{gilmore09}
{Gilmore}, R.~C., {Madau}, P., {Primack}, J.~R., {Somerville}, R.~S., \&
  {Haardt}, F. 2009, \mnras, 399, 1694

\bibitem[{{Gu} {et~al.}(2001){Gu}, {Cao}, \& {Jiang}}]{gu01}
{Gu}, M., {Cao}, X., \& {Jiang}, D.~R. 2001, \mnras, 327, 1111

\bibitem[{{Jorstad} {et~al.}(2005)}]{jorstad05}
{Jorstad}, S.~G. {et~al.} 2005, \aj, 130, 1418

\bibitem[{{Jorstad} {et~al.}(2010)}]{jorstad10}
---. 2010, \apj, submitted, arXiv:1003.4293

\bibitem[{{M{\"u}cke} \& {Protheroe}(2001)}]{muecke01}
{M{\"u}cke}, A. \& {Protheroe}, R.~J. 2001, Astroparticle Physics, 15, 121

\bibitem[{{Murray} \& {Chiang}(1995)}]{murray95}
{Murray}, N. \& {Chiang}, J. 1995, \apjl, 454, L105+

\bibitem[{{Reimer}(2007)}]{reimer07}
{Reimer}, A. 2007, \apj, 665, 1023

\bibitem[{{Shakura} \& {Sunyaev}(1973)}]{shakura73}
{Shakura}, N.~I. \& {Sunyaev}, R.~A. 1973, \aap, 24, 337

\bibitem[{{Sikora} {et~al.}(2008){Sikora}, {Moderski}, \&
  {Madejski}}]{sikora08}
{Sikora}, M., {Moderski}, R., \& {Madejski}, G.~M. 2008, \apj, 675, 71

\bibitem[{{Stecker} {et~al.}(2006){Stecker}, {Malkan}, \& {Scully}}]{stecker06}
{Stecker}, F.~W., {Malkan}, M.~A., \& {Scully}, S.~T. 2006, \apj, 648, 774

\bibitem[{{Tosti} {et~al.}(2008){Tosti}, {Chiang}, {Lott}, {Do Couto E Silva},
  {Grove}, \& {Thayer}}]{tosti08_atel}
{Tosti}, G., {Chiang}, J., {Lott}, B., {Do Couto E Silva}, E., {Grove}, J.~E.,
  \& {Thayer}, J.~G. 2008, The Astronomer's Telegram, 1628, 1

\bibitem[{{Vercellone} {et~al.}(2009)}]{vercellone09}
{Vercellone}, S. {et~al.} 2009, \apj, 690, 1018

\bibitem[{{Vercellone} {et~al.}(2010)}]{vercellone10}
---. 2010, \apj, 712, 405

\end{thebibliography}
\end{document}